\documentclass[12pt,preprint]{aastex}

\usepackage[top=1in, bottom=1in, left=1in, right=1in]{geometry} 

\usepackage{amsmath,amssymb,latexsym, graphicx}

\begin{document}
\setlength{\parskip}{0pt}

\shorttitle{Necessity of Acceleration-Induced Nonlocality}
\title{Necessity of Acceleration-Induced Nonlocality}

\author{Bahram Mashhoon\altaffilmark{*}}
\altaffiltext{*} {mashhoonb@missouri.edu}

\affil{Department of Physics and Astronomy,
       University of Missouri, Columbia, MO 65211, USA}
  
\

{\hspace*{6.3cm}
\textbf{Abstract} }
\\

The purpose of this paper is to explain clearly why nonlocality must be an essential part of the theory of relativity. In the standard local version of this theory, Lorentz invariance is extended to accelerated observers by assuming that they are pointwise inertial. This locality postulate is exact when dealing with phenomena involving classical point particles and rays of radiation, but breaks down for electromagnetic fields, as field properties in general cannot be measured instantaneously.  The problem is corrected in nonlocal relativity by supplementing the locality postulate with a certain average over the past world line of the observer.

\section{Introduction\label{intro}}

The principle of relativity---namely, the assertion that the laws of microphysics are the same in all inertial frames of reference---refers to the measurements of ideal inertial observers. The transition from Galilean invariance of Newtonian physics to Lorentz invariance marks the beginning of modern relativity theory.  We thus assume throughout that Lorentz invariance is a fundamental symmetry of nature. The basic laws of microphysics have been formulated with respect to ideal inertial observers, since these are conceived to be free of the various limitations associated with actual observers.  At this point a basic dichotomy is encountered involving theory and experiment: while the basic laws of non-gravitational physics have all been formulated with respect to ideal inertial observers, the experimental basis of these laws---namely, the foundation of physical science---has been established via actual observers that are all more or less accelerated.

Ideal inertial observers are imaginary and do not really exist. They have been introduced so as to embody the principle of inertia perfectly. Each is supposed to move with constant speed on a straight line from minus infinity to plus infinity. There is in fact no way to verify by experiment that an observer moves uniformly forever from the infinite past to the infinite future, since distant past and future states of the universe are not directly accessible to experimentation.
Moreover, repeated attempts to determine that an object is in fact moving uniformly on a rectilinear path will sufficiently disturb its motion such that the net momentum transfer in the measurement process will render the motion nonuniform. The nature and extent of such disturbances can be investigated in order to develop velocity meters of high accuracy.
  
Simply stated, the fundamental microphysical laws, such as the principles of quantum mechanics, have been formulated for nonexistent ideal inertial observers, while all actual observers are accelerated. The resolution of this dichotomy requires an a priori axiom that relates inertial and accelerated observers. The observational consequences of such an axiom should then be compared with experimental results. 

The standard theory of relativity is based on the assumption that an accelerated observer is pointwise inertial. Thus the association between actual accelerated observers and ideal inertial observers is purely \textit{local} in the theories of special and general relativity [1]. This local relation has been generalized to a nonlocal one in recent years [2-9]. The main goal of the present paper is to explain in a conceptually transparent manner the necessity of this nonlocal extension of the standard relativity theory. To this end, we imagine a global background inertial frame of reference, with coordinates  
$x^{\mu}=(ct,\mathbf{x})$, as the arena for our considerations. According to the standard local theory, what a pointwise-inertial accelerated observer in this arena would measure in  a physics experiment is determined by connecting the momentary inertial frames along its world line with the background global frame by pointwise Lorentz transformations. 

What do accelerated observers really measure? As this is the main issue under discussion here, it is necessary to discuss our theoretical treatment of observers. The term ``observer" is employed here in an extended sense; that is, observers can be sentient beings or measuring devices. As such, observers are extended systems in space; however, following the standard practice in the theory of relativity, we represent an observer by a single world line for the sake of simplicity. This is \textit{not} considered to be a fundamental limitation; rather, it helps simplify the analysis. Furthermore, an actual measuring device may have various limitations due to its mode of construction or its mode of operation. For instance, it may break down and fail to operate properly under certain conditions. In practice, all these issues have to be carefully considered; however, for the purposes of this theoretical discussion, we assume that all observers are free of such limitations and are in this sense ideal. Henceforth, we concentrate on the theoretical distinction between ideal inertial observers and ideal accelerated observers. 

\section{Locality}

The locality postulate states that an accelerated observer is at each instant physically equivalent to an otherwise identical momentarily comoving inertial observer. The latter follows the straight world line that is tangent to the world line of the accelerated observer at that instant. Thus an accelerated observer may be replaced in effect by an infinite sequence of hypothetical momentarily comoving inertial observers; mathematically, the world line of the accelerated observer is the envelope of the straight world lines of the corresponding hypothetical inertial observers. 

This locality assumption originates from Newtonian mechanics, where the state of a point particle is determined at each instant by its position and velocity.  The point particle and the hypothetical comoving inertial particle of the same mass share the same state and are thus physically equivalent.  Consider, for instance, the motion of a Newtonian point particle of mass $m$ under an external force $\mathbf{f}(t,\mathbf{x},\mathbf{v})$ in the background global inertial frame. According to Newton's laws of motion, 
\begin{eqnarray}
\frac{d\mathbf{x}}{dt}=\mathbf{v}, && \frac{d\mathbf{v}}{dt}=\frac{1}{m}\mathbf{f}(t,\mathbf{x},\mathbf{v})~~.
\end{eqnarray}
\\
At any given time $t$, the specification of the particle state $(\mathbf{x},\mathbf{v})$ uniquely determines the motion for all  time. Moreover, if $\mathbf{f}$ is turned off at any time $t$, the motion continues uniformly on a straight line tangent to the path at $t$. The inertial motion of the particle is thus continually interrupted by the action of the external force that changes the state of the particle. This is the physical explanation for the fact that the path of the particle under the external force is the envelope of all such tangent lines. 

It follows that the postulate of locality is automatically satisfied in the Newtonian mechanics of point particles, as it is ingrained in the Newtonian laws of motion.  Minkowski formally extended the locality postulate in a natural way to relativistic physics [10].

\section{Limitations of Locality}

Relativity theory grew out of developments in electrodynamics. Therefore, in addition to the relativistic mechanics of point particles, traditional relativistic physics involves classical electromagnetic fields and radiation. The question naturally arises whether the locality postulate can be extended to fields. We are therefore interested in the field measurements of ideal accelerated observers. In general, such measurements cannot be performed instantaneously.

It seems intuitively clear that an accelerated observer may be considered practically inertial during an experiment if the observer's acceleration is such that its motion is in effect uniform and rectilinear for the duration of the physical process under study. Lorentz invariance can then be employed to predict the result of the experiment.  Let ${\lambda}/{c}$ be the intrinsic timescale for the process under consideration, and let $L/c$ be the acceleration timescale over which the velocity of the observer changes appreciably; then, the condition for the validity of the connection between theory and experiment is

\begin{eqnarray}
\lambda\ll L~~.
\end{eqnarray}

Consider, for example, the measurement of the frequency of an incident plane monochromatic wave of frequency  $\omega$ and wave vector $\mathbf{k}$, $\omega=c\left|\mathbf{k}\right|$, by an accelerated observer. If the observer moving with velocity 
$\mathbf{v}(t)$ could be considered inertial, then the local inertial frame of the observer can be related to the background global inertial frame by a Lorentz transformation. Subsequently, the Doppler effect may be employed to give 

\begin{equation}
\omega^{'}(t)=\gamma\left[~\omega-\mathbf{v}(t)\cdot\mathbf{k}~\right]~~.
\end{equation}
\\
Physically, the observer must register at least a few oscillations of the incident wave before an adequate determination of its frequency can be attempted; on the other hand, Eq. (3) is valid if during this time $\mathbf{v}(t)$ changes very little.  We can express this condition from the standpoint of the fundamental inertial observers---that is, those at rest in the global background inertial frame---as

\begin{eqnarray}
nT\left\vert\frac{d\mathbf{v}(t)}{dt}\right\vert\ll v(t)~~.
\end{eqnarray}
\\
Here $T=2\pi/\omega $ is the period of the incident wave, $v(t)$ is the magnitude of $\mathbf{v}(t)$ and $n$ is the number of cycles of oscillation needed for a reasonable determination of the frequency. Thus $n>1$ is an integer. 

Let us first assume that the observer is linearly accelerated with acceleration $\mathbf{a}=d\mathbf{v}/dt$; using 
$\omega=2\pi c/\lambda$ and $v(t)<c$, Eq. (4) can be written as 

\begin{eqnarray}
\lambda\ll\frac{c^{2}}{a}~~,
\end{eqnarray}
\\
where $a$ is the magnitude of the acceleration of the observer. 
If, on the other hand, the observer is at the time rotating uniformly on a circular orbit with frequency $\Omega$, we have $a=\Omega~v$. Then, Eq. (4) implies that $nT\Omega\ll 1$ or

\begin{eqnarray}
\lambda\ll\frac{c}{\Omega}~~.
\end{eqnarray}
\\
Equations (5) and (6) illustrate Eq. (2) in these specific instances. 

In general an observer has a translational acceleration length $L=c^{2}/a$ and a rotational acceleration length $L=c/\Omega$---see [11, 12] for formal definitions of these fundamental quantities. For observers at rest on the surface of the Earth, $c^{2}/g_{\oplus}\approx 1$ light year and $c/\Omega_{\oplus}\approx 28$ astronomical units.  These are very large compared to wavelengths of interest in the laboratory; therefore, for most physical situations the acceleration of the observer can be neglected for the duration of the physical process under consideration. This explains why locality is so effective in practice. 

The acceleration lengths ($c^{2}/a$ and $c/\Omega$) are familiar concepts in standard relativistic physics; they indicate the spatial limitations associated with accelerated coordinate systems. However, these lengths emerge from the basic heuristic argument presented here with a different fundamental significance: Lorentz invariance can be directly extended to accelerated observers only when $L\gg\lambda$. The significance of this conceptual analysis is the establishment of the notion that the standard theory of relativity is valid so long as intrinsic wave phenomena are considered in the eikonal (JWKB) approximation. Furthermore,  the locality postulate is an essential ingredient of Einstein's heuristic principle of equivalence; therefore, similar restrictions are expected in the general theory of relativity [11-15]. 

To have a complete theory, it is therefore necessary to go beyond the locality postulate, since the standard local theory of relativity is not capable of dealing with phenomena involving $\lambda\gtrsim L$. To illustrate this point, consider the path of an accelerated charged particle. The radiation emitted by the particle has dominant wavelengths $\lambda\sim L$; therefore, locality is violated for the particle. That is, the accelerated particle's measurements cannot be theoretically predicted by pointwise Lorentz transformations from the background inertial frame to the momentarily comoving inertial frame at the position of the particle along its path. On the other hand, the particle radiates away its energy via electromagnetic waves and this loss of energy must show up in its equation of motion through a radiation reaction force. In the nonrelativistic approximation, the Abraham-Lorentz equation of motion of the particle of charge $q$ is given by

\begin{eqnarray}
\frac{d\mathbf{v}}{dt}-\frac{2}{3}\frac{q^{2}}{mc^{3}}\frac{d^{2}\mathbf{v}}{dt^{2}}+ \cdot\cdot\cdot=\frac{1}{m}\mathbf{f}(t,\mathbf{x},\mathbf{v})~~.
\end{eqnarray}
\\
The Newtonian postulate of locality no longer holds here as $\mathbf{x}(t)$ and $\mathbf{v}(t)$ are not sufficient to specify the state of the particle at time $t$; for this purpose, at least the acceleration of the charged particle is needed as well. This violation of locality is of course due to the interaction of the particle with the electromagnetic field. 

These considerations have an interesting consequence: If the physical processes of interest all involve 
$\textit{pointlike coincidences}$ involving point particles and rays of radiation such that in effect $\lambda=0$, then Eq. (2) is automatically satisfied and the accelerated observer may be considered $\textit{pointwise}$ inertial. That is, at each instant the observer's measurements depend only upon its position and velocity, but not upon its acceleration. This is indeed the locality postulate of the standard theory of relativity. 

To prepare the ground for a basic modification of the locality postulate, we must first ensure that the ideal accelerated observers have access to $\textit{standard}$ measuring devices. These are defined to operate in accordance with the locality postulate. To see clearly why they are needed here, imagine an observer in a laboratory fixed on the Earth trying to determine the influence of the laboratory's acceleration on the measurement of a certain electromagnetic field. A series of measurements extending over the past world tube of the laboratory are needed for the eventual determination of the field. If the separate measurements in the series are not carried out using ideal $\textit{standard}$ measuring devices, the end result of the experiment would depend on the influence of acceleration on the inner workings of the devices employed.

\section{Standard Measuring Devices}

Ideal measuring devices that are so robust as to be essentially unaffected by acceleration are called ``standard". Historically, they were first invoked in the theory of relativity in connection with the so-called ``clock hypothesis". 

Consider fundamental inertial observers with inertial coordinates $x^{{'}\mu}=(ct^{'},\mathbf{x^{'}})$ at rest in a frame moving with constant velocity $\mathbf{v}$ with respect to the background frame. It follows from the invariance of spacetime interval under Lorentz transformations that

\begin{eqnarray}
dt^{'}=\left(1-\frac{v^{2}}{c^{2}}\right)^{1/2}dt~~.
\end{eqnarray}
\\
Since by the locality postulate an accelerated observer passes through an infinite sequence of momentarily equivalent inertial observers, its local proper time $\tau$ is a sum of infinitesimal time intervals each of the form of Eq. (8). Therefore, the proper time is given by

\begin{eqnarray}
\tau=\int\left[1-\frac{v^{2}(t)}{c^{2}}\right]^{1/2}dt~~.
\end{eqnarray}
\\
For an ideal accelerated clock, of course, various internal inertial effects exist; however, by definition, the clock that registers $\tau$ is \textit{standard}. The assertion that such clocks actually exist is the content of the clock hypothesis. Indeed, ideal standard rods and clocks are routinely employed in relativity for theoretical discussions of spacetime measurements. 

From a modern perspective, all ideal measuring devices that are pointwise inertial are standard. That is, an ideal measuring device is practically standard if we can suppose that over the length and time scales characteristic of typical measurements the net impact of the internal inertial effects over the operation of the device can be neglected. \textit{The main point of this paper is that even when an accelerated observer employs only ideal standard devices for measurement purposes, there are intrinsically nonlocal measurements involving electromagnetic fields that extend over the past world line of the observer and
hence go beyond the postulate of locality}. 

From a mathematical standpoint, the locality postulate in effect replaces the world line of the observer at each instant by its tangent at that event. Geometrically, the tangent line is the first Frenet approximation to the curve. The Frenet-Serret method of moving frames for world lines has been discussed in Ref. [16]. Each fundamental inertial observer is naturally endowed with an orthonormal tetrad frame that consists of the four unit basis vectors of the Lorentz frame in which the observer is at rest.  The locality postulate therefore implies that an accelerated observer carries an orthonormal tetrad frame $\lambda^{\mu}_{~(\alpha)}(\tau)$ such that at each instant of proper time, this frame coincides with that of the momentarily comoving inertial observer. It follows from the method of moving frames that 

\begin{eqnarray}
\frac{d\lambda^{\mu}_{~(\alpha)}}{d\tau}=\Phi_{(\alpha)}^{~~(\beta)}(\tau)~\lambda^{\mu}_{~(\beta)}~~,
\end{eqnarray}
\\
where $\Phi_{{(\alpha)}{(\beta)}}$ is the antisymmetric acceleration tensor of the observer. 

It is a direct consequence of the locality postulate that the projection of various tensorial quantities on the orthonormal tetrad frame of the accelerated observer can be physically interpreted as the measurement of such quantities by the observer.  For instance, given an electromagnetic field $F_{\mu\nu}(x)$ in the background global inertial frame, 

\begin{eqnarray}
F_{(\alpha)(\beta)}(\tau)=F_{\mu\nu}\lambda^{\mu}_{~(\alpha)}\lambda^{\nu}_{~(\beta)}
\end{eqnarray}
\\
is the field measured by the hypothetical comoving inertial observer at $\tau$, which is also what the accelerated observer would measure according to the standard local theory.

The establishment of the local tetrad frame is ultimately based on the standard clock and measuring rod that the accelerated observer may use for local spacetime determinations. To treat the fundamental problem of field measurement in the next section, we may tentatively assume the existence of ideal standard devices and return to a deeper examination of this issue once the nonlocal theory has been properly formulated.

\section{Nonlocality}

The thought experiment of Sec. 3 in connection with the measurement of the frequency of an electromagnetic wave involves a process that is not instantaneous, so that the observer along its world line needs to measure the radiation field for some time before a determination of its frequency becomes possible; therefore, as a matter of principle, a certain integration of data over the past world line of the observer is necessarily involved in the measurement process. 
   
A further significant step in the analysis of the measurement process involving an accelerated observer is the recognition that the electromagnetic field itself cannot be measured instantaneously. This general assertion, which applies to any electromagnetic field, is the content of the Bohr-Rosenfeld principle [17, 18].  Bohr and Rosenfeld pointed out that though Maxwell's equations involve $\mathbf{E}(t,\mathbf{x})$ and $\mathbf{B}(t,\mathbf{x})$, only spacetime averages of these fields have immediate physical significance. They arrived at this conclusion by analyzing the process of field measurement by the fundamental inertial observers on the basis of the Lorentz force law. Once the necessity of such averaging is established for ideal inertial observers, it becomes clear that it must be extended to ideal accelerated observers. It is indeed essential in this case due to the existence of the intrinsic acceleration scales. Approximating the world tube of the accelerated observer by a world line, we conclude in accordance with causality that it is necessary to take the past world line of the accelerated observer into account in any field determination. 
  
To search for a proper generalization of the local connection between inertial and accelerated observers, we return to the relationship between the accelerated observer and the infinite sequence of momentarily comoving inertial observers along its world line.  The averaging process in the Bohr-Rosenfeld principle is linear; hence, we look for a general linear connection between the field as determined by the accelerated observer, $\mathcal{F}_{(\alpha)(\beta)}(\tau)$,  and the  field  determinations  $F_{(\alpha)(\beta)}(\tau)$ of the momentarily  comoving inertial observers.  The most general relation of this type is given by

\begin{eqnarray}
\mathcal{F}_{(\alpha)(\beta)}(\tau)=F_{(\alpha)(\beta)}(\tau)+u(\tau-\tau_{0})\int_{\tau_{0}}^{\tau}K_{(\alpha)(\beta)}^{~~~~~(\gamma)(\delta)}(\tau,\tau^{'})
F_{(\gamma)(\delta)}(\tau^{'})d\tau^{'}~~.
\end{eqnarray}
\\
Here $\tau_{0}$ is the instant at which the acceleration is turned on and $u(t)$ is the unit step function such that $u(t)=0$ for $t<0$ and $u(t)=1$ for $t>0$. The kernel $K$ is expected to be directly proportional to the acceleration of the observer; therefore, Eq. (12) expresses the sum of two terms that contribute to the measured field: the local connection together with an ``average" over the past world line of the accelerated observer.  That is, the locality postulate by itself implies pointwise equivalence, $\mathcal{F}_{(\alpha)(\beta)}(\tau)=F_{(\alpha)(\beta)}(\tau)$, while the nonlocal averaging involves a weight function $K$ that vanishes in the absence of acceleration. Linearity and causality thus lead to Eq. (12); the problem then reduces to the determination of the kernel $K$. 
   
The basic extension of the locality postulate in Eq. (12) can be naturally expressed for any field. We emphasize that ansatz (12) is envisioned to be in accordance with measurements performed with \textit{standard} devices; therefore, the nonlocal term in this ansatz is expected to be independent of any measuring device employed and is thus purely induced by the acceleration of the observer in Minkowski spacetime. The situation here is reminiscent of the correspondence between wave mechanics and classical mechanics.  The linear memory of past acceleration is then a vacuum effect. One may contemplate further extension of these ideas; for instance, Eq. (12) may have to be generalized to a nonlinear relation in the presence of a medium. 
   
As the inertial observers along the path are fictitious, one can interpret Eq. (12) as a formal expression for the field as measured by the accelerated observer at $\tau$ given in terms of the projections of the field $F_{\mu\nu}(x)$ on the observer's tetrad frame along its past world line. The upshot of this discussion is that the locality postulate must be amended by the memory of past acceleration.  After the acceleration has been turned off, the memory effect persists; in fact, it turns out to be a constant field in nonlocal special relativity [3]. It is important to emphasize that nonlocal special relativity actually involves only local fields that satisfy certain integro-differential field equations carrying the memory of past acceleration.

\section{Spin-Rotation Coupling}

The main physical notion that we employ to constrain the nonlocal kernel is derived from a peculiar consequence of the locality postulate as applied to the general phenomenon of spin-rotation coupling. 
   
Imagine an observer that rotates uniformly with frequency $\Omega$ about the direction of incidence of a circularly-polarized plane monochromatic electromagnetic wave of frequency $\omega$. What is the wave frequency $\omega^{'}$ as measured by the rotating observer? Projecting the incident field on the observer's natural tetrad system and Fourier analyzing the result, namely, $F_{(\alpha)(\beta)}(\tau)$, we find

\begin{eqnarray}
\omega^{'}=\gamma(\omega\mp s\Omega)~~,
\end{eqnarray}
\\
where $s=1$ is the spin of the field and the upper (lower) sign refers to an incident positive (negative) helicity wave. Unlike the frequency, the measured amplitude is independent of helicity, however.  These general results follow from pointwise field determination in accordance with the locality postulate.   
   
Equation (13) involves the transverse Doppler effect as well as helicity-rotation coupling and has been discussed extensively in previous papers [4, 19, 20]. The helicity-rotation part has a simple intuitive explanation: In a positive (negative) helicity wave, the electric and magnetic fields rotate in the positive (negative) sense with the wave frequency 
$\omega$ about the direction of propagation; therefore, the rotating observer perceives the rotation frequency of the field in the incident wave  to be $\omega-\Omega$ $(\omega+\Omega)$ instead. For $\Omega\ll\omega$ and $\gamma\approx 1$, Eq. (13) has been verified via the GPS with $\omega/(2\pi)\sim 1$ GHz and $\Omega/(2\pi)\sim 8$ Hz [21]. The spin-rotation coupling is a manifestation of the inertia of intrinsic spin and for $\omega\gg\Omega$ has been the subject of a number of investigations---see [22-28] and the references cited therein.  
   
For $\omega=\Omega$ in the case of incident positive-helicity radiation in Eq. (13), $\omega^{'}=0$. Let us note that $\lambda=2\pi L$ in this case, so that locality is expected to fail here. By a mere rotation, the whole radiation field stands completely still with respect to the class of all observers uniformly rotating with frequency $\omega$ about the direction of incidence. The radiation field loses its temporal dependence, but is sinusoidal in space. This situation is entirely analogous to a consequence of the pre-relativistic Doppler formula when an inertial observer could move along the direction of wave propagation with the speed of light. Einstein noted this defect in his autobiographical notes---see page 53 of Ref. [29]. Just as Lorentz invariance avoids the problem for $\textit{inertial}$ observers---that is, an inertial observer cannot stay at rest with an electromagnetic wave---nonlocal relativity theory should be constructed in such a way that $\textit{no observer}$ can ever stay at rest with an electromagnetic wave.  Indeed, this is the main postulate that is employed in nonlocal special relativity for the determination of the kernel. That is, the nonlocal kernel is so chosen as to correct a perceived defect in the standard local special relativity theory. 
   
In this work, we have emphasized classical electromagnetic fields; however, our treatment can be naturally extended to the quantum domain. For a Dirac field, for instance, Eq. (13) holds with $s=1/2$. Clearly, the basic dichotomy between theory (involving ideal inertial observers) and experiment (involving actual accelerated observers) that was presented in Sec. 1 applies equally well to quantum theory and to preserve wave-particle duality and Heisenberg's uncertainty principle, it is necessary to insist that the Dirac field should not lose its temporal dependence with respect to any observer [30]. 
   
The phenomenon of spin-rotation coupling has been employed to develop nonlocal kernels for the basic physical fields [31-33]. The special case of the electromagnetic field has been treated in Ref. [34]. The main nonlocal kernels together with the essential formal elements of nonlocal special relativity have been presented in Ref. [3]. The theory is in agreement with all observational data available at present. 
 
It is interesting to point out some of the observational consequences of the resulting nonlocal special relativity for the spin-rotation coupling under consideration in this section. Expressing the incident radiation in terms of its electromagnetic vector potential, a relation analogous to Eq. (12) is obtained for the vector potential as determined by the rotating observer---see, for instance, [4]. When this quantity is Fourier analyzed, we recover Eq. (13) except when $\omega=\Omega$ in the positive-helicity case. In this resonance situation, the nonlocal contribution to the vector potential is linearly dependent upon proper time. For the wave amplitude, we find that nonlocality renders it helicity dependent. This can be heuristically seen from 

\begin{eqnarray}
\int e^{-i(\omega\mp s\Omega)t}dt=i\frac{e^{-i(\omega\mp s\Omega)t}}{(\omega\mp s\Omega)}~~,
\end{eqnarray}
\\
since the kernel turns out to be constant for uniformly accelerated observers. Thus for the observationally accessible regime $\omega\gg\Omega$, the amplitude is larger (smaller) when the helicity of the incident particle is in the same (opposite) sense as the rotation of the observer. 
   
These purely nonlocal predictions of the theory in the case of spin-rotation coupling for normal incidence agree qualitatively with quantum mechanical results in the correspondence regime [35]. Using Bohr's correspondence principle, the nonrelativistic orbital motion of electrons in the limit of large quantum numbers can be compared with the motion of rotating observers. A detailed investigation reveals that the predictions of the nonlocal theory have closely related counterparts in quantum mechanics; therefore, nonlocal relativity is in better agreement with quantum theory than the standard theory of relativity [35]. These encouraging results notwithstanding, it is important to subject nonlocal special relativity to direct experimental tests [3, 4].

\section{Discussion}

The physics of accelerated systems in Minkowski spacetime has been critically examined in this paper. For sufficiently low accelerations, the relevant acceleration scales can be large enough to ensure that $L\gg\lambda$, in which case the standard local theory is quite adequate. Otherwise, the local theory breaks down. We have then delineated the conceptual steps that must be taken to rectify the situation and explained clearly the theoretical necessity of a nonlocal generalization of the standard theory of relativity. An essentially complete nonlocal theory of special relativity together with some of its observational consequences has been outlined in Refs. [3, 4] and awaits direct confrontation with observation. 

The equivalence principle indicates a deep connection between inertia and gravitation. When nonlocality is thus extended to the gravitational domain, a nonlocal generalization of Einstein's general theory of relativity emerges in which nonlocal gravity simulates dark matter [7-9].
\\
\\
\\
{\hspace*{6.6cm}

\clearpage

{\hspace*{6.1cm}
\textbf{References} }
\\

{
[1] A. Einstein, \textit{The Meaning of Relativity} (Princeton University Press, Princeton, NJ, 
     \hspace*{2cm}1955).

[2] B. Mashhoon, \textit{Nonlocal theory of accelerated observers}, Phys. Rev. A \textbf{47}, 
      4498  \hspace*{2cm}(1993). 

[3] B. Mashhoon, \textit{Nonlocal special relativity}, Ann. Phys. (Berlin) \textbf{17}, 11705 (2008).

[4] B. Mashhoon, \textit{Optics of rotating systems}, Phys. Rev. A \textbf{79}, 062111 (2009).

[5] J. Ramos and B. Mashhoon, Phys. Rev. D \textbf{73}, 084003 (2006).

[6] B. Mashhoon, Ann. Phys. (Berlin) \textbf{16}, 57 (2007).

[7] F. W. Hehl and B. Mashhoon, Phys. Lett. B \textbf{673}, 279 (2009).
       
[8] F. W. Hehl and B. Mashhoon, Phys. Rev. D \textbf{79}, 064028 (2009).
       
[9] H.-J. Blome, C. Chicone, F. W. Hehl and B. Mashhoon, Phys. Rev. D \textbf{81}, 065020  \hspace*{2cm}(2010).

[10] H. Minkowski, in \textit{The Principle of Relativity}, by H.A. Lorentz, A. Einstein,
H. \hspace*{2cm}Minkowski and H. Weyl (Dover, New York, 1952).

[11] B. Mashhoon, Phys. Lett. A \textbf{143}, 176 (1990).
      
[12] B. Mashhoon, Phys. Lett. A \textbf{145}, 147 (1990).
       
[13] B. Mashhoon, in \textit{Relativity in Rotating Frames}, edited by G. Rizzi 
   and M.L. \hspace*{2cm}Ruggiero (Kluwer Academic Publishers, Dordrecht, 2003), pp. 43-55.
      
[14] B. Mashhoon and U. Muench, Ann. Phys. (Berlin) \textbf{11}, 532 (2002).
      
[15] B. Mashhoon, Lect. Notes Phys. \textbf{514}, 211 (2003).

[16] J. L. Synge, \textit{Relativity: The General Theory} (North-Holland, Amsterdam, 1971).

[17] N. Bohr and L. Rosenfeld, K. Dan. Vidensk. Selsk. Mat. Fys. Medd. \textbf{12}, \hspace*{2cm}No. 8  (1933); translated in \textit{Quantum Theory and Measurement}, edited by  J.A. \hspace*{2cm}Wheeler and W.H. Zurek (Princeton University Press, Princeton, NJ, 1983).
      
[18] N. Bohr and L. Rosenfeld, Phys. Rev. \textbf{78}, 794 (1950).

[19] B. Mashhoon, Gen. Rel. Grav. \textbf{31} (Hehl Festschrift), 681 (1999).

[20] J. C. Hauck and B. Mashhoon, Ann. Phys. (Berlin) \textbf{12}, 275 (2003).

[21] N. Ashby, Living Rev. Relativity \textbf{6}, 1 (2003).

[22] B. Mashhoon, Phys. Rev. Lett. \textbf{61}, 2639 (1988).

[23] F.W. Hehl and W.-T. Ni, Phys. Rev. D \textbf{42}, 2045 (1990).

[24] B. Mashhoon, Phys. Lett. A \textbf{198}, 9 (1995).

[25] B. Mashhoon, R. Neutze, M. Hannam and G.E. Stedman, 
         Phys. Lett. A \textbf{249},  \hspace*{2cm}161 (1998).
        
[26] G. Papini, Phys. Rev. D \textbf{65}, 077901 (2002).
         
[27] B. Mashhoon and H. Kaiser, Physica B \textbf{385-386}, 1381 (2006).

[28] K. Y. Bliokh, J. Opt. A: Pure Appl. Opt. \textbf{11}, 094009 (2009).

[29] P. A. Schilpp, \textit{A. Einstein: Philosopher-Scientist} (Library of Living Philosophers,  \hspace*{2cm}Evanston, Illinois, 1949).

[30] B. Mashhoon, Phys. Rev. A \textbf{75}, 042112 (2007).

[31] C. Chicone and B. Mashhoon, Ann. Phys. (Berlin) \textbf{11}, 309 (2002).
        
[32] C. Chicone and B. Mashhoon, Phys. Lett. A \textbf{298}, 229 (2002).

[33] F. W. Hehl and Y.N. Obukhov, \textit{Foundations of Classical Electrodynamics} (Birkh$\ddot{\mathrm{a}}$user,  \hspace*{2cm}Boston, 2003).

[34] B. Mashhoon, Phys. Lett. A \textbf{366}, 545 (2007).

[35] B. Mashhoon, Phys. Rev. A \textbf{72}, 052105 (2005). 
}

\end{document}